\documentstyle[12pt]{article}
\title  { Symmetries of the relativistic two-boson system in external field.}

\author { Philippe Droz-Vincent\\[2mm]LUTH\\
Meudon \footnote{Observatoire de Paris, CNRS, Universit\'e Paris Diderot,
5  place Jules Janssen,   92195  Meudon, France   }}

\date {}   

\newcommand {\zer}{ {(0)} }

\newcommand {\Del}{\Delta}

\newcommand{\calp}{ {\cal P} }

\newcommand{\beq}{\begin{equation}}
\newcommand{\eeq}{\end{equation}}

\newcommand{\Gam}{\Gamma}
\newcommand{\GGam}{ { \scriptscriptstyle \Gamma} }
\newcommand {\DDel}{ {\scriptscriptstyle \Delta} }
\newcommand {\AAA}{  {\scriptscriptstyle A}   }
\newcommand {\BBB}{  {\scriptscriptstyle B}   }

\newcommand{\ome}{\omega}
\newcommand{\taub}{ {\overline \tau} } 

\newcommand{\gam}{\gamma}
 \newcommand {\half}{ {1 \over 2}}
 \newcommand {\noi}{\noindent}

 \newcommand{\mun}{{\mu \nu}}

\newcommand {\ttt}{  {\scriptscriptstyle T} }
\newcommand {\lll}{  {\scriptscriptstyle L} }
\newcommand {\del}{\delta}

\newcommand {\alp}{\alpha}

\newcommand{\Zhat}{ {\widehat    Z} }
\newcommand  {\disp}{\displaystyle}

  \newtheorem{prop}{Proposition}
 \newtheorem{theo}{Theorem}

\newcommand{\beprop}{\begin{prop}}
\newcommand{\betheo}{\begin{theo}}
\newcommand{\enprop}{\end{prop}}
\newcommand{\entheo}{\end{theo}}
\begin{document}
\maketitle
\abstract{
We investigate the survival of   symmetries in   a  relativistic system of 
 two mutually interacting bosons  coupled with an external field, when this field  is    "strongly" translation invariant  in some directions  and  additionally   remains  unchanged  by  other isometries  of  spacetime.  
Since the relativistic interactions cannot be composed additively, it is not {\em a priori}
garanteed that  the  two-body system inherits all the symmetries of the  external potential.
However, using  an  ansatz which  permits to preserve   the compatibility  of the mass-shell
 constraints  in  the presence of the  field, we  show  how  the "surviving isometries" 
can actually be implemented  in the  two-body  wave equations. 
}

$$ \           $$

  \bigskip

\section {Introduction, Notation}

\noi   Applying an external field to a particle  generally  spoils Poincar\'e invariance. 
But in  many  cases of interest  some piece of this invariance  still survives,
 because the external field 
itself  exhibits  certain  kind of symmetry; for instance a static Coulomb field applied to a charged particle preserves spherical symmetry though it breaks space translation invariance.  

\noi
At least   insofar as scalar particles are concerned,   the symmetries of  the field could be characterized as the symmetries  of  the  one-body 
motion in this field because (through Noether's theorem)  they  are  automatically reflected  in  the motion of a  test particle. 

When external forces are applied to a  system of {\em  several } particles  undergoing mutual interactions, it is tempting to expect a similar situation;
in other words it would be natural to formulate  a   general principle of   invariance  under the  surviving isometries, as follows:

\noi {\sl   Principle of  Isometric Invariance} 

\noi   {\em  If the external potentials  applied to the system remain invariant under a transformation of the Poincar\'e group, then the  system should enjoy the same symmetry}.

The  Galilean analog of this statement is  trivial, 
because  usually all  the  interactions arise  additively in the  non-relativistic  Hamiltonian. 

\noi  Insofar as the equation of motion is concerned, the relativistic dynamics of a {\em single} particle
 automatically agrees with the principle~\cite{salamanca}. In contrast,  as soon as $N >1$,  it is by no means  obvious  that 
$N$-body  {\em relativistic} dynamics  can  always   be  constructed in agreement with the
   principle of isometric invariance.

\noi  Indeed  relativistic  interactions cannot  be just  linearly  composed; such a  complication is bound to arise in  any  formulation of  relativistic dynamics (see for instance  the  work of  Sokolov~\cite{soko}  using  the  "point form" of dynamics).

\noi The main goal of this article consists in proving that, given a system of two  mutually interacting particles, the coupling of this system to a large class of external fields  can be  actually realized  in a way  that  satisfies this principle.

 For  analyzing these matters there exist many formulations of  relativistic particle dynamics,
but   the  more appropriate ones are those which  make  use of manifestly covariant mass-shell   constraints~\cite{tod}~\cite{conf}. In this framework the motion is generated by the (half)squared-mass operators  and  is  governed by  a system of $N$ coupled wave equations~\cite{DV}~\cite{etc} 
In the two-body case, the relationship between this approach and the conventional methods of quantum field theory has been established~\cite{saz}~\cite{bijetal}. 
An advantage of the  constraint formalism over the Bethe-Salpeter equation is the natural elimination of the relative-time degree of freedom.  Let us rather  emphasize that in the context of mass-shell constraints (which admits a classical analog with  Poisson brackets in a phase space) symmetries and first integrals have a clear-cut status: for example a constant of the motion is characterized by  its  commutation with both squared-mass operators.

For simplicity we focus on the case of two {\em scalar particles} which interact between themselves and  are also  submitted to  external forces.
Assuming that we  explicitly know the  term describing mutual interaction alone, 
 the first problem is  to write down   wave equations that  remain  compatible when the  external  field   is applied to the system; another requirement is obviously in order: one must retrieve the correct limits when either the mutual interaction or the external field vanishes.

\noi 
In general this problem is not tractable in closed form, and the necessary requirements stated above are not sufficient for a full determination of the wave equations.  Complementary information must be obtained either from the underlying quantum field theory or    
from  reasonable assumptions  of "simplicity" which  would  actually  involve some  implicit  symmetry.
The principle  of isometric invariance provides a natural prescription for removing or at least reducing the ambiguities.

\noi
We shall concentrate on the cases where the external field  is translation invariant (in a special way referred to as {\em strong}) along some directions of spacetime,  because this situation  allows for mass-shell constraints  in closed form.

\noi 
A first solution  was  given by J. Bijtebier~\cite{bij} under the hypothesis that   the  applied field is strongly stationary  along a (implicitly unique) timelike direction. 
 We put forward a more systematic  formulation  which only requires  that the external potential  is  strongly translation invariant 
 along one {\em or several} directions of spacetime \cite{droz}. Such directions are labelled as "longitudinal" and, in the generic case,    their orthogonal complement is spanned by
  the "transverse" ones; in this study we exclude the  exceptional  case  where the longitudinal directions span a null manifold. 
This approach  provides  an Ansatz  which permits to  explicitly  write  down  the squared-mass operators  in a new  representation; these operators in turn are strongly  translation invariant,  implying  that  strong-translation  invariance is  automatically  preserved  by  the  coupling .

\noi     
Naturally,  beside   strong translation invariance, it may happen that  the external potential remains  unchanged  also under  some other  isometries.
 
 \noi For instance (in suitable coordinates) a constant magnetic field  not only is strongly translation invariant along the directions that span the plane  (03) but also exhibits rotational  symmetries   in the  planes (12)  and  (03).
 
\noi  The above principle  would require that also these extra symmetries are preserved in the motion of {\em two}  charged particles, even  when  we take their mutual interaction into account. 
In this situation  the question  arises as to know  whether the squared-mass operators  furnished by the ansatz   actually respect  these additional symmetries.
 
  Although we mainly have in mind the case of a constant magnetic field, we present here a general treatment valid for any  
 external field which enjoys  strong translation invariance.  
  Note that up to now the merit of the ansatz  was   to  provide  squared-mass operators  that  reduce to the correct limits when any of the interactions vanishes.  But  the ansatz will appear  more  satisfactory  if  we  further prove that it respects  isometric invariance.

\medskip

\noi In order to tackle this question we are thus led to consider  the (continuous)  isometries  of  spacetime that survive  as   symmetries  of the system  in the presence   of an  external  field.
    
Section 2 deals with  one-body motion in external fields admitting directions of strong translation invariance. In Section 3, after a brief sketch of the two-body problem in general,   we focus on the case of two independent particles submitted  only  to  external fields; their symmetries and invariances are discussed.   Mutual interactions are introduced in Section 4, and  concluding remarks are reserved to Section 5.

\noi Greek indices take on the values $0, 1, 2, 3 $.

\section{Symmetries  in the one-body motion}
We  consider the potential  created by the field, {\em i.e.}
 the interaction term, $G(q, p)$   which arises in the   single-particle
 Hamiltonian equation of   motion  $2K \psi =    m^2 \psi$.
The half-squared mass operator is
\beq  K =  \half  p^2  +  G        \label{KG1}    \eeq
For instance for  the charge $e$  in an electromagnetic field, using the Lorentz gauge and the canonical commutation relations
$ \disp    [q^\alp , p_ \beta ] =  i  \del ^\alp _\beta             $
we  have
\beq     2G =  -e A \cdot p  -  e p \cdot  A  +  e^2  A \cdot  A   \eeq
Similarly, in a weak  gravitational field such that  
$  \disp   g^{\mu \nu}  =  \eta ^{\mu \nu}  +  h ^{\mu \nu}  (x)   $      we would have
$ 2G =         p_ \mu    h ^{\mu \nu}    p _ \nu  $.

\medskip

In general, any quantity which commutes with $K$ is a constant of the motion.
Any quantity which commutes with $G$  canonically generates a  transformation which leaves the  external potential invariant. Because of the physical importance of linear and angular momenta we focus on the canonical transformations that correspond to the continuous {\sl isometries of spacetime}    (displacements).

\noi  The presence of $G$ breaks  the  full Poincar\'{e} invariance. But it may happen that some  element of the      Poincar\'{e}  Lie    algebra $\calp$   still commutes with $G$. Let $j$ be {\em any} element of  $\calp$, we call it  a {\sl momentum}  and    we   may write
\beq        j =  a  ^\alp   p_\alp  +  \ome ^\mun    m_ \mun      
\label{defj}    \eeq
where   $m_ \mun =  q _ \mu p_\nu    -  q _ \nu  p_ \mu  $,  
for  some  constant vector    $ a  ^\alp$  and some  constant  skew-symmetric tensor    $   \ome  ^\mun$.  
This  terminology  encompasses linear  and  angular  momentum.

\noi
Since $p^2$ is a Casimir  of   $\calp$  
 it is clear that $[K , j]$
 vanishes (and $j$ is a constant of the single-particle motion)  iff  
        $  [G , j ] = 0 $.

\noi   In this case   $j$ is a  {\sl conserved momentum 
in the one-body sector}.  

\noi $j$ generates a canonical transformation referred to as a {\sl  surviving isometry}.
\medskip
Among all the surviving isometries there may be  some translations:
 {\em  $G (q, p )$  is  simply  translation  invariant  along direction  $w^\alp$
 when  $  [G ,   w  \cdot   p ]$  vanishes}.
 But  among the  symmetries   respected by  the presence of  $G$  we shall distinguish  {\sl strong translation invariance} defined as follows:

\noi { \em  $G$ is   strongly translation invariant along  direction  $w^\alp$  when
both  $[G ,   w  \cdot  q   ]$   and   $  [G ,  w  \cdot   p ]$  vanish }.

\noi For instance  if  $a ^\mu  =  (1, 0, 0, 0 )$, we say  that $G$ is strongly stationary along direction $a$  when both  
 $[G ,   q^0 ]$   and   $  [G ,   p^0 ]$  vanish,  etc.

\noi This notion is basically defined within the one-body sector, although it will be  useful  essentially  in two-body problems. Note also  that  strong translation invariance  can  be   already  considered at the  classical level,  in  terms of  Poisson brackets   in  the  eight-dimensional  one-body  phase space.

\medskip   
\noi
The directions of strong translation invariance span the {\sl longitudinal  space} $E ^\lll$.
Assuming that  $E ^\lll$ admits orthonormal frames  (this case will be referred to as "generic" in contrast to the exceptional case  where $E ^\lll$ is a null plane)  we introduce the {\sl transverse} space
  $E ^\ttt$ as its  orthocomplement. So the space of four-vectors is an orthogonal direct sum
  \beq   E =   E ^\lll \oplus    E ^\ttt  \label{split}   \eeq
  In terms of  the   projector  onto   $ E ^\lll$,
say   $\tau  ^\alp _ { \    \beta} $,  we  distinguish  longitudinal and transverse parts  of the
canonical variables, say 
$ q_{a \lll} ^\alp ,  p_{b \lll } ^\beta$   and  
  $ q_{a \ttt} ^\alp ,  p_{b \ttt } ^\beta$ respectively.  More generally   
 we   define purely longitudinal (resp. transverse ) quantities.
   
\noi  The Lie algebra of the Poincar\'e group gets split  along the same line and we have a
longitudinal subalgebra   $\calp _\lll$  generated by   
$ \tau ^{\mu  \alp}    p _ \alp$ 
  and   $\tau ^{\mu   \alp}  \     \tau ^{\nu  \beta}  \      m_ {\alp \beta} $.
It is  obvious that  any element of    $\calp _\lll$  remains  a  conserved   momentum and  generates a  surviving   isometry.
But 
it  may  happen  that  other isometries  also  survive the application of external field.
$$ \                     $$
\medskip

\noi {\bf Example}
 
\noi
Consider  a  charge  $e$ submitted to a constant electromagnetic field $F_{\mun}$
 such that only    $F_{12}= - F_{21} =F  \not= 0$.
The interaction term in the Hamiltonian equation of motion is
\beq  G = - {e \over 2}     (q_1 p_2  - q_2 p_1 )  F
 -  {e^2 \over 8} ( (q_1)^2 +   (q_2 )^2 ) F ^2
                                              \label{magconst}    \eeq
This system is {\em strongly}  translation invariant along any direction of the 
two-dimensional plane $(03)$.  

\noi We may equally observe that  it  is   invariant not only by rotation in this plane, but  also  by rotation in the plane $(12)$ (the latter   generated by  the  transverse  angular  momentum     $m_{12}$).

\noi Another constant of the motion is the pseudo-momentum
$ C = p+ e A$,  but  its conservation results from invariance under
the so-called "twisted translations"  that {\em are not}  spacetime isometries~\cite{simon}. 
\bigskip

\noi   In general, in the presence of strong translation invariance it is convenient  to classify  all the conserved isometries.
To this end we    split any four-vector $\xi$ as  
$ \xi ^ \mu  = ( \xi ^\AAA , \xi ^ \GGam ) $  
 where  Latin (resp. Greek) capitals refer to the  longitudinal (resp. transverse) directions. 
In this notation the  longitudinal and transverse parts of  the canonical coordonates are 
$$ q^\mu _\lll  =  ( q^\AAA , 0)  , \qquad   q^\nu _\ttt  =  ( 0,  q^\GGam ) ,
   \qquad  \quad 
  p _{ \lll \nu } =  ( p_\AAA , 0) , \qquad    
p _{\ttt \nu}  =  ( 0,  p _\GGam )  $$
For an arbitrary momentum $j$ like in  (\ref{defj}) the skew-symmetric tensor  $\ome ^\mun$ can be written as

\beq    \ome  ^\mun  =  \     \left(    \begin{array}{cc}
  \ome ^{\AAA \BBB}    &              \ome ^{\AAA  \DDel}  \\
\ome ^{\GGam  \BBB }  &                \ome ^{\GGam  \DDel}
                                 \end{array}    \right)          \label{splitome}     \eeq
where   of  course   $ \ome ^{\GGam  \BBB }  =   -   \ome ^{\BBB  \GGam}    $.
 We get
$$ \ome ^\mun q_\mu p_\nu  =      \ome ^{\AAA  \BBB}  q_\AAA   p_\BBB  + 
   \ome ^{\AAA  \GGam} q_\AAA   p_\GGam     +
   \ome ^{ \GGam  \BBB}  q_\GGam   p_\BBB      +
   \ome ^{\GGam  \DDel}  q_\GGam   p_\DDel                                  $$                                            
and so on.   We cast   (\ref{defj})  into the form of a unique decomposition
\beq    j =     j_{(\lll )}    +     j_{(\ttt )}     +     j_{\rm  mix} 
                                                     \label{decompoj}   \eeq
where  
\beq  j_{(\lll )} =        a^\AAA  p_\AAA   +  2 \ome ^{\AAA \BBB}  q_\AAA p_\BBB
 \label{newdefjl}      \eeq
   
\beq  j_{(\ttt )} =        a^\GGam  p_\GGam   +
   2 \ome ^{\GGam \DDel}  q_\GGam p_\DDel       \label{newdefjt}      \eeq

\beq   j_{\rm  mix}  =    2 \ome ^{\AAA \DDel}  q_\AAA  p_\DDel    +
  2 \ome ^{\DDel \AAA }  q_ \DDel p_\AAA     \label{newdefjmix}      \eeq

\noi  Any operator  which involves only $q_\lll$ and  $p_\lll$
(resp. $q_\ttt$ and  $p_\ttt$)    is  called     {\sl  longitudinal} 
(resp. {\sl transverse}).
Beware that a  longitudinal {\em component} of a vector is  not necessarily a longitudinal {\em operator} . 

\noi In particular we can consider  longitudinal and transverse momenta; for instance  
$\disp   \ome   ^{\AAA  \BBB}  m_{\AAA \BBB}  $ is a longitudinal  rotation, etc. 
The splitting   (\ref{split})  determines, in  $\calp$
two remarkable subalgebras   namely $\calp _\lll$  and  $\calp_\ttt$
 formed by the longitudinal and  transverse  momenta  respectively.


\medskip
\noi   As noticed previously,  
 
 \noi {\sl  Any  longitudinal momentum is a constant of the motion} 
  
  \noi although every longitudinal  momentum  is not necessarily   the  generator  of   a  longitudinal translation. 
Therefore  insofar as conservation is concerned  the  nontrivial  piece,  in  formula 
 (\ref{decompoj}) above,     is  the reduced quantity
\beq  j_{\rm red}      =  j_{( \ttt )}  +  j_{\rm mix }      \eeq
It is clear that $j$  survives  as a constant of the motion   iff   $j_{\rm red}$    does.  
Since it belongs to    $\calp$  
it commutes with  $p^2$, 
thus according to (\ref{KG1})  it commutes with  $K$  iff
$$  [j_{\rm red}  , G  ] = 0 $$
In view of  (\ref{newdefjt})(\ref{newdefjmix})
 we get on the one hand
\beq            [j_{(\ttt)}  , G ] =
   a ^\GGam   [ p_{ \GGam} ,\    G ]      + 
 2 \ome  ^{\GGam  \DDel} \      [q _ \GGam    p_{ \DDel} , \     G ]  
\label{newjtG}   \eeq
Since  $G$ is purely transverse, neither $q_\AAA$ nor  $p_\BBB$
can  arise in the expression  of $ [j_{(\ttt)}  , G ] $.    

\noi  On the other hand we  derive  from  equation     (\ref{newdefjmix})   
\beq     [j_{\rm  mix}  , G ] =
 2 \ome   ^{\AAA  \DDel}  \       q_ { \AAA }  [p_{ \DDel} ,  G ]
+ 2\ome ^{\DDel   \AAA}  \     [q_ {  \DDel}  ,\    G]  p_ {\AAA}  
\label{newjmixG}   \eeq 

\noi But     $ [q_ {  \GGam}  ,\    G]  $  and  also 
     $ [p_ {  \DDel}  ,\    G] $   
   are  purely transverse;  it  follows  that $q_{  \AAA}$   and 
    $p_{ \BBB}$    arise  only  linearly in
 this expression,  so   $ [j_{\rm  mix}  , G ]$ is simply linear  {\em and  homogeneous} with respect to  the  longitudinal canonical variables. Thus in order to have  
$$     [j_{(\ttt)}  , G ] =   -    [j_{\rm  mix}  , G ]   $$ 
 both sides of this  formula  must  vanish,  which  amounts to  have  
  both      $ j_{(\ttt)} $   and  $ j_{\rm  mix} $  separately   conserved.  

\noi    This  situation  is  expressed  by  the conditions 
\beq            a ^\GGam   [ p_{ \GGam} ,\    G ]      + 
 2 \ome  ^{\GGam  \DDel} \      [q _ \GGam    p_{ \DDel} , \     G ]  =0 
\label{newcondisum}   \eeq
\beq  \ome ^{\AAA \GGam}  \    [p_\GGam , \   G]  = 0  
\label{condimixp}     \eeq
\beq  \ome  ^{\GGam \AAA}      \    [q_\GGam , \   G]  = 0  
\label{condimixq}     \eeq

\noi    Taking into account the antisymmetry of $\ome$ it is clear that  the last two formulas imply the following:

\noi   keep  the label   $A$  fixed  and  consider the vector  
$w ^\mu  =  (0 ,  w^\GGam  =  \ome ^{\AAA \GGam}  ) $.
 Then  the quantities  
$ w \cdot  q$  and  $w  \cdot  p$  commute with $G$, in other words $w$ is a  direction of strong translation invariance  (unless it  vanishes).
But  $w$ being purely  transverse this would  clash with the very definition of   $E^\ttt$    
(which states that {\em all} such  directions are  included in  $E^\ttt$).    And this for all  $A$.
Thus  all the mixed components  
$  \ome ^{\AAA \GGam} $   must  vanish, and  no  $j_{\rm  mix}$  can  be  a conserved  momentum. 
In other words 
\betheo
No mixed momentum can be a constant of the motion in external field.
\entheo

\noi  {\bf Corollary  1} $\qquad  $    {\em Any conserved momentum takes on the form 
$ j =     j_{(\lll)}    +   j_{(\ttt)}   $,  where 
          $j_{(\lll)} $     and   $ j_{(\ttt)}   $  are  separately  conserved}.

\medskip
\noi
 {\bf Example}: for a  constant magnetic external field,  with only $F_{12} \not=  0$ 
  
\noi   we have  
$ {\scriptstyle A,   B  }=  0,3 $ 
 whereas  $ {\scriptstyle \Gam ,  \Del}  =  1, 2$.
$$   q_\lll = ( q^0 ,0,0, q^3 )  ,  \qquad        
\quad     q_\ttt = (0,  q^1 , q^2 , 0 )    $$
and so on.

\noi   $\calp_\lll$   is  spanned by     $p_0 , p_3 , m_ {03}$ whereas       $\calp_\ttt$   is  spanned by   
   $p_1 , p_2 , m_ {12}$.
        These  Lie algebras respectively  obey  the formulas
\beq   [p_0 ,p_3 ] =0, \qquad  \
 [p_0 , m_{03} ]  =    -i p_3  , \qquad
            [ p_3 , m_{03}   ]  =  -i p_0                                 \label{lielongi}               \eeq
\beq   [p_1 , p_2 ] =0, \qquad  \
 [p_1 , m_{12} ]  =  i p_2    , \qquad
            [ p_3 , m_{12}   ]  =    -i  p_1                             \label{lietrans}               \eeq

\noi  The purely transverse quantity 
$ j_{(\ttt)} =   m_{12} $  remains  conserved. 

\noi For this example  we can directly  check that 
no  mixed  momentum  can  survive:  
if it  were so,   condition   (\ref{condimixq})
would be satisfied for  some  choice  of  the  coefficients 
  $\ome ^ {\Gam  \Del}$.  Since the   splitting  of  spacetime directions   is  $2\oplus 2$,    there are at most  four independent such coefficients,
say    $\ome ^{10} , \ome ^{13} ,  \ome ^{20} , \ome ^{23}$.
From  (\ref{magconst}) we  derive
$$ [q_1 , G ] =   - {e \over 2}  F_{12} \     q_2  ,    \qquad  \quad    
   [q_2 , G ] =     {e \over 2}  F_{12} \     q_1                         $$
Inserting into     (\ref{condimixq})    yields
$$  \ome^{1  A}  \     q_2    -  
    \ome^{2  A}  \     q_1  = 0  $$
 But  the  transverse canonical coordinate $q^1 , q^2 $  are independent, therefore 
   $ \ome^{1  A }  $   and     $   \ome^{2  A} $
must vanish for  both 
 ${\scriptstyle A} = 0$  and      ${\scriptstyle A} =   3$.
Finally  the four components of   $\ome ^ {1 A }$  and     $\ome ^ {2 A }$  are zero,  which excludes the possibility that a nonvanishing  $j_{\rm mix }$  be  conserved.

\section { Two-body motion}

In the two-body sector the canonical  variables   are $q_a , p_b$  submitted to 
the commutation relations
$$        [q_a ^\mu ,  p_{b \nu}   ] = i \    \del _{ab}  \     \del ^ \mu _\nu     $$ 
with  $a,b, c   =   1, 2$.
We separate the relative variables according to
$$ z^\alp =  q_1 ^\alp - q_2 ^\alp ,   \qquad \quad
       y^\beta =   \half (p_1 ^\beta    -  p_2 ^\beta  )    $$
It is convenient to set
$$ Z =    z^2  P^2  -   (z \cdot P )^2     $$

\noi
 The Poincar\'{e}  Lie algebra is  realized  in terms  of  
the generators~\footnote{
In the  formulas concerning the two-body sector,  $1,2$ are particle labels.
In contradistinction, in   formulas    (\ref{magconst})   and
(\ref{lietrans})      devoted to the single particle,  the  indices $1,2$ obviously   refer to    spacetime directions.}

\beq   P = p_1 + p_2        ,\qquad  \  
       M= (q_1  \wedge p_1   ) _\mun  +   (q_2  \wedge p_2   ) _\mun 
                                                      \label{defpoinc}      \eeq
We can consider individual  momenta  $j_1 , j_2$  where $j_a $ depends only on $q_a , p_a $ and  set
 \beq   J = j_1  + j_2                \label{Jcollec}           \eeq
 so that the  generator of  any  spacetime isometry  takes on the form
\beq  J =    a  ^\alp   P_\alp  +   \ome ^\mun    M_ \mun       \label{defJ}    \eeq   
In the case of two independent  ({\em  i.e.}  not mutually interacting)   particles, 
the square-mass operators  are  $2K_1 , 2K_2$  with 
 $K_a = K (q_a , p_ a ) =  \half  p_a ^2   + G_a$.

\noi
When  (in addition to  external coupling)   the particles   are mutually interacting,
the   individual   variables   cannot  any  more   appear separately   in  the  equations of motion.
The  square-mass operators   are  generally   written  as
$2H_1 , 2H_2$   and
\beq  H_a  =  K_a   + V                        \eeq
where  the  interaction  term  $V$    depends on  the canonical coordinates of {\em both} particles. $V$    must  be  chosen with  care, such that   
$[H_1 , H_2 ]$ vanishes  and  such that    Poincar\'{e}   invariance is restored  in  the limit where the 
external field is turned off.
 We define  $V^\zer$  as the  no-external-field  limit  of  $V$ (more generally the label $(0)$ refers to an isolated system).   Naturally $V^\zer$ is supposed to  commute with 
{\em  all} the generators of   spacetime isometries.    In other words,  in the absence of external field, 
 $H_1 , \      H_2 $   respectively   reduce  to    $H_1 ^\zer  , \      H_2  ^\zer $  where
$ \disp  H_a ^\zer  =  \half  p_a  ^2   +   V^\zer   $.
In practice   $    V^\zer   $   is  explicitly given (as a    Poincar\'{e}   invariant  operator) and  is  such  that
     $H_1 ^\zer  $  commutes  with  $    H_2   ^\zer $.
For a large class of  mutual interactions we can  write
\beq    V^ \zer =   f (Z , P^2 ,  y \cdot  P  )     
                          \label{Vzero}          \eeq 
Realistic forms of $f$  have been derived from quantum field theory~\cite{cratvan}\cite{sazjallou}.  
  
\noi As  soon as one  assumes the presence of an external field,  one has to modify  $    V^\zer   $      in such a way  
  that     now       $H_1 $    commutes  with    $  H_2 $.    
The problem is   of course  nonlinear  and  an    explicit  solution  is   available only   for special classes  of  external
potentials.  
This solution  is not  completely unique:  further considerations are needed in order to remove (at least partially) the   arbitrariness. For this purpose isometric invariance will  be a criterium of choice.

\subsection{Two  independent particles in external fields}

{\em In the limit where no mutual interaction is present},  the two-body motion is fully determined by the external potentials.  A   spacetime  infinitesimal isometry  (generated by  $J$) is a   symmetry
of the system {\em  as a whole}  when  both  $G_1$  and  $G_2$  commute with its generator, say
 $$ [J, G_1 ] = [J , G_2 ]  =  0                                    $$

\noi   
This isometry  is  a  {\sl surviving isometry}.

\noi     In this case $J$ is a  first integral  for the motion of  two  independent particles    respectively submitted to the potentials  $G_1 , G_2 $. 
In view of  (\ref{Jcollec}) it is clear that  any  momentum 
$J$   survives   as  a constant of the two-body motion   iff  
\beq 
 [j_1 , G_1 ]  =  [j_2 , G_2 ] = 0                   \label{0}
\eeq
in other words  $j_1 $  and   $j_2 $   respectively survive  application of  the  external potentials  $G_1$ and $G_2$ in the one-body problem.

 \noi Surviving isometries  may include rotations and translations. 
A translation  along $w$ is a surviving isometry provided that 
$w \cdot P$ commutes with both potentials, which makes  $w$ at least a direction of {\em simple} translation invariance in the two-body sector, say
\beq  [G_a , w \cdot P ] =0     \eeq
or equivalently
\beq  [G_1 , w \cdot  p_1 ] =   [G_2 , w \cdot  p_2 ] =   0   \eeq
But we shall be more specially interested  by  {\sl strong  translation} invariance, defined as follows  by analogy  with the one-body  case; we  say that

\noi {\bf \sl  Definition}$\    $  {\em   The {\em couple of potentials} $G_1 , G_2$ is strongly translation invariant along direction $w$  when   each  potential  separately  is strongly  translation invariant  along  $w$ in the one-body sector},
 in other words
\beq                                                             
[ G_1 ,  w \cdot q_1 ]  =   [ G_1 ,  w \cdot p_1 ]  = 0,  \qquad   \     
 [ G_2 ,  w \cdot q_2 ]  =  [ G_2 ,  w \cdot p_2 ]  = 0, 
\eeq  

\noi    When  they  exist, the directions of strong translation invariance (for  the   two-body sector) span    a linear subspace   $E^\lll$
  included in the space of four-vectors,  and  the  projection of  any  vector  onto  $E^\lll$  is obtained with help of a  tensor $\tau$.

\medskip

For distinguishable particles it may happen that  
 $G_1$ and $G_2$ be strongly invariant along {\em distinct} longitudinal spaces, $E_1 ^L , E_2 ^L$ (this situation would correspond to the existence of two distinct projectors  $\tau _1 , \tau _2$). Still the {\em common} directions of strong translation invariance span the linear space 
$E ^\lll  = E_1 ^\lll  \cap  E_2 ^\lll$  corresponding  to  a  single   projector $\tau$.
\noi But in general  $ E_1 ^\lll  ,   E_2 ^\lll$  and  $E ^\lll$ might  be all differents, which would  allow  the possibility of  different splittings  in the one-body sector and the two-body one.

For simplicity we shall focus  on the simple case where  
 $ E_1 ^\lll  =   E_2 ^\lll  =  E ^\lll$.
This situation is ensured by assuming  that the external couplings are of the same kind for both particles, in the following sense:

\noi   {\bf \sl Definition} $\     $ {\em The  external  couplings are {\em of the same kind for both particles}  when there exists a one-body potential $G (\alp ,q , p )$ where $\alp$ is a coupling parameter,  such that 
$\disp  G_a  =   G(\alp _a ,q_a , p_a ) $ for  $a = 1,2$,  with  nonvanishing  coupling constants  $\alp _1 , \alp_2$.}

\noi  In other words both are submitted to the same field with possibly distinct coupling constants. The most simple example  is  given by  two different charges  if we neglect their mutual interaction in front of  the  external field.
Our definition  discards the special case where one coupling constant, say $\alp _2$, vanishes because (if $\alp _1 \not= 0$) it leads to  $E_1 ^\lll  \not=  E_2  ^\lll$.

\medskip

 To summarize, a surviving isometry may involve rotations and translations, the latter being strong or not.
    In the sequel we shall assume  that
    
 \noi    a) {\em  the  external potential   admits  one or several directions of strong translation invariance} 

 \noi   b) {\em  both couplings are of the same kind and  $E ^\lll$ is generic (not a null plane)}.
  
\noi     Again the space of four-vectors is split   as  in  (\ref{split})   and  we define  the   longitudinal piece of any vector, say 
$ \xi  ^\alp  _\lll  =  \tau  ^\alp _\beta  \xi ^ \beta   $.  Similarly we separate the longitudinal canonical variables  
$ \disp  \     
   q_{a \lll} ^\alp   =  \tau ^ \alp  _  {\beta}       q^\beta _a  $
and 
$\disp    p_ {b  \lll} ^\beta  =  \tau  ^\beta _ \gam   p^\gam _{b  }                        \   $  
from the  transverse ones, say
$$  q_{a \lll} ^\alp   =  (q_a  ^\AAA  ,  0 )  ,  \qquad  
    q_{a \ttt} ^\alp   =  ( 0 , q_a  ^\GGam )  ,  \qquad  
    p_ {b  \lll} ^\beta  = (p_ b ^\BBB  , 0 )  ,  \qquad   
    p_ {b  \ttt} ^\beta  = ( 0 , p_ b  ^ \DDel  )        $$

\noi  Since    $ E_1 ^\lll  =   E_2 ^\lll  =  E ^\lll$ we have  
\beq
  J_{(\lll}) =  j_{1 (\lll )}  +  j_ {2  (\lll )}    ,        \qquad  \qquad   
    J_{(\ttt}) =  j_{1 (\ttt )}  +  j_ {2  (\ttt )}         \label{A}
\eeq

\noi  The  external  potentials    $  G_a $  are purely transverse operators since they are supposed to commute with  $q_{a \lll}$ and  $p_{b \lll}$.

\noi  With help of  $\tau$  and  
  $\taub ^\alp _ \beta    =  \del ^\alp _\beta   -  \tau   ^\alp  _\beta $
      we define  
$ \ome    ^{\AAA  \BBB} $,  etc, as in (\ref{splitome}).

\noi In the two-body sector, the Lie algebra  
 of the  Poincar\'e  group,  say    $\   ^2 \calp$             has  the generators $P_\rho$  and $M_\mun$. In view of  (\ref{split}) it has   longitudinal and  transverse subalgebras,
  say    $\   ^2 \calp_\lll$   and   $\   ^2 \calp_\ttt$ respectively.
 For instance in the constant magnetic case    $\   ^2 \calp_\lll$ is  spanned  by
$P_0 , P_3 ,  M_{03}$, with commutators  analogous to those in formula (\ref{lielongi}).

\noi Any element of  the  Poincar\'e   algebra  
 can be written as in  (\ref{defJ}), but also
\beq  J=    a \cdot  P  + 
 \ome ^{\AAA  \BBB}  M_{\AAA \BBB}   +
  \ome ^{\GGam \DDel}  M_{\GGam \DDel}    +
    \ome ^{\AAA  \GGam}   M  _   {\AAA  \GGam}    +
   \ome  ^{ \GGam  \BBB}   M  _   { \GGam   \BBB}   
                               \label{redefJ}          \eeq   
if we split  the tensor  $\ome$  into four pieces corresponding to 
purely longitudinal (resp. transverse ) parts and the mixed parts, say
  $ \ome _{\AAA \BBB} ,       \ome_{\GGam  \DDel}  ,
       \ome _ {\AAA \DDel} ,    \ome _{ \GGam   \BBB}     $.
We get 
\beq  J =  J_{(\lll )} +  J_ {(\ttt )}   +   J _ {\rm mix} 
                 \label{decompo}  \eeq
with  
\beq  J_{(\lll )} =  a ^ \AAA   P_\AAA   +
     \ome  ^ {\AAA   \BBB}    M _{\AAA  \BBB}   \label{defJL}      \eeq

\beq     J_{(\ttt )}  =  a ^\GGam      P_ \GGam    +
     \ome  ^{\GGam   \DDel }     M _{\GGam     \DDel}   \label{defJT}        \eeq
\beq     J _ {\rm mix}   =   
  \ome   ^ {\AAA  \GGam }   M _ {\AAA  \GGam }          +
   \ome  ^{\DDel   \BBB}   M_ {\DDel  \BBB}  
=   2  \ome  ^ {\AAA  \GGam }   M _ {\AAA  \GGam }
                         \label{defJmix}              \eeq
A glance  at    (\ref{newdefjmix})   shows  that  
\beq   J_{\rm  mix}  =      j_{1 {\rm  mix}}  +   j_{2 {\rm  mix} }
                                                            \label{B}       \eeq    
 
Now, in search for the conditions which make a momentum $J$ to be conserved in the motion of two independent particles, looking at formulas
(\ref{0}) (\ref{A}) (\ref{decompo}) (\ref{B}) we are left with two separate problems in the one-body sector.   Applying the results of the previous Section we obtain  an extension  of  Theorem 1  and Corollary 1 to the two-body sector,

\beprop       For independent particles no $J _{\rm  mix} $  can be conserved,  and when  $J$ is conserved we have  
$J_{(\lll )}$         and   $ J_{(\ttt )} $   separately conserved.
\enprop


\noi  So any isometry  of  spacetime can be decomposed as in (\ref{decompo}).
The first piece is  $\disp  J _{( \lll )}$
which  depends only on  $q ^ \AAA  $ and  $p_{\BBB }$  thus  commutes with  $G_1 ,  G_2$. In other words 

\beprop    All  purely longitudinal momenta  survive  as  constants  of the motion of two independent particles.
\enprop

\noi In contradistinction the purely {\em transverse} momenta may fail to be conserved.
  For instance in the magnetic example above, $P_1 , P_2$
 are not conserved although $M_{12}$  is.




\section{Mutually interacting particles}

\noi  The  Ansatz is as follows~\cite{bij}~\cite{droz}.
The   external-field representation  is   formally   obtained  with help of
 $ {\rm e}  ^{iB}$ where $B= TL$  is  the commutative  product of a  transverse  operator by a  longitudinal one, namely
\beq    T =  y_ \ttt    \cdot   P _ \ttt  +  G_1  -  G_2             \label{defT}          \eeq 
\beq    L  =  {   P_ \lll  \cdot  z_\lll    \over       P_ \lll  ^2 }  \label{defL}    \eeq
The transformed square-mass operators are  
  \beq  H '_a  =  K'_a   + V'         \label{Haprim}                      \eeq 
with 
\beq  K' _1 + K' _2  =  K_1 + K_2    - 2 T  { y_\lll  \cdot  P _\lll    \over  P_\lll  ^2}
                                        + {T^2   \over        P_\lll  ^2}               \label{somKprim}            \eeq
\beq            K' _1   -   K' _2   =  y_      \lll    \cdot   P _ \lll         \label{difKprim}       \eeq 
\beq    V'  =  f ( \Zhat , P^2 ,  y_\lll  \cdot P_\lll  )                               \label{satz}   \eeq
where  $f$  is  the  function  in   (\ref{Vzero}) 
and   $\disp   \Zhat     =  {\rm e} ^{ib}  Z  {\rm e} ^{-ib}$  
where  $b$  is  the  no-field limit  of  $B  = L T $. Namely
\beq    \Zhat   =   Z   +   2   (   z_\ttt  \cdot  P   ) ( z_\lll     \cdot  P       )    -
           (  z_\lll   \cdot  P  )  ^2  \              {P_\ttt  ^2   \over   P_\lll  ^2  }   \label{defZhat}         \eeq

\medskip
\noi  {\sl Remark   \    }
     The formulas    (\ref{somKprim}) -    (\ref{defZhat})                              describe the {\em external-field representation}. In the absence of external field  this representation   reduces  to the usual one  {\em only after a unitary transformation}.
Indeed, according to    (\ref{defT}), we  see that $T$ and  $B$  are  not 
cancelled  by the vanishing of  $G_1 , G_2$.  

\medskip

\noi  The following statement is trivial, 
and can be checked by hand using the canonical commutation relations,

 \beprop
The quantities 
$\qquad  y_\ttt \cdot  P_\ttt ,  \qquad  
 z_\ttt  \cdot  P  =     z _\ttt  \cdot   P _\ttt  ,
  \qquad  P_\ttt ^2  \quad  $ 
are  invariant under the  transverse isometries, in other words they commute with every $J_{(\ttt )}$.

\noi          The quantities
$ z_\lll  \cdot  P  =     z _\lll  \cdot   P _\lll ,  \qquad
       y_\lll  \cdot  P_\lll , \qquad       P_\lll ^2  \quad  $
are  invariant by    the   longitudinal  isometries, in other words
they commute with every     $J_{(\lll )}$.  
\enprop

\noi In addition the  transverse quantities commute with all $J_{(\lll )}$ 
and         the  longitudinal quantities commute with all $J_{(\ttt )}$. 


\noi
{\bf Corollary  2   \   }   {\em  Any   $J_{(\lll )}   +   J_{(\ttt )} $ commutes  with  $\Zhat$}  
(irrespective of    $ [J_\ttt ,  K_a ]$  vanishing  or  not).



\beprop   If a  momentum  $J$   survives as a constant of the motion of independent particles, it is not affected by the transformation generated by  $B$.   
\enprop

\noi  {\sl Proof} 

\noi  From Proposition 1 we know that such a momentum is
  $  J = J_{(\lll )}   +   J_{(\ttt )} $
where both     $ J_{(\lll )}$  and      $ J_{(\ttt )}$ 
  commute  with  $K_1 ,  K_2$. So  all we have to prove is that the change of representation  generated  by  $LT$ produces  $J' =J$.

\noi   So first consider   $ J_{(\lll )}$, it obviously commutes with $T$. To prove that it commutes with $B$  we just have to check that it also commutes with $L$,  but  in  (\ref{defL})  it is manifest 
 that  $L$  is  invariant  by the  longitudinal displacements, 
  in  other words we have  
 $ [ J_{(\lll )} ,   L ] = 0$, thus      $ [ J_{(\lll )} ,   B ]$   vanishes.

\noi Now consider   $ J_{(\ttt )}$, being purely transverse it commutes with $L$.  
Still we are concerned about    $ [ J_{(\ttt )}   ,  T ]$ where  $T$ is as in  (\ref{defT}). In $T$ the first   term     
   $y_ \ttt    \cdot   P _ \ttt  $   is manifestly  invariant   by  all  translations   and  invariant by     the   transverse   rotation,  thus  
 $y_ \ttt    \cdot   P _ \ttt  $  commutes with    $J _{( \ttt )}$.     
The second term  in  $T$  is   $G_1 - G_2$, 
but    $ J_{(\ttt )}$  is  supposed to  commute     with   $K_1$  and  $K_2$,  hence  also  with   $G_1 $  and   $G_2$   and  finally  with  $T$.  
To summarize  $ [B, J ]  $ vanishes which  implies  that       $J' = J$.   []


\bigskip

\betheo
In the context of the ansatz, with  both external  couplings  of the same  kind, if a   momentum $J$  is a constant of the motion of two independent particles, it remains a constant of the motion in the presence of mutual interaction.
\entheo

\noi Proof

\noi  We want to prove that    $[H' _a  , J' ] $ is  zero.
Our  assumptions  mean   that   $ [   K_a  ,  J] =0 $    or  equivalently that 
$ [ K' _a  ,  J' ] = 0  $.  But Propo 4 implies that  $J' = J $, so we have that
  $ [ K' _a  ,  J ] = 0 $.  In view of    (\ref{Haprim}) all we have to check now is  whether in  (\ref{satz})  all the infredients of  $V'$  actually commute with $J$.
Propo 1  tells that  $J =  J _ {(\lll )} +  J _ {(\ttt )}$. Corollary 2  ensures that  $\Zhat$  commutes with $J$.  Propo 3 implies that $P^2$  and  $ y_\lll \cdot  P_\lll$  have the same property, which achieves the proof. []

\medskip

As an example consider two charges in a constant magnetic field:
the momenta  $P_0 , P_3, M_{03} ,  M_{12} $  remain conserved  in the presence of a mutual interaction defined as in   (\ref{Vzero}).



\section{Summary and conclusion}

We have proposed a principle of invariance  which seems to be a natural requirement  in the presence of external fields.
Then we focused on the case of  external fields admitting strong translation invariance.
In a first step  we  checked that, in the absence of mutual interaction,  the description  obtained in the one-body sector  can  be re-phrased  with the same structure in the two-body  framework
 (at this stage all  surviving isometries are easily identified  and each  one  obviously  corresponds to a conserved  momentum).
   Then  we have introduced  the  mutual coupling, assuming that  the  composition of all the interactions together is performed according to the Ansatz.
  And finally we have shown that, in the generic case  and provided  both  external  couplings are of the same kind,  this procedure ensures that the spacetime isometries which leave the external potential invariant remain symmetries of the two-body system submitted to all the interactions.
So  the  interacting two-body system  inherits the conservation laws implied  by   the  spacetime invariances of the external field.
To summarize,  the principle of isometric invariance is satisfied at least in the context  of  strong translations, and this result 
 enhances our confidence in the ansatz.

 \noi In the present  paper we took the view that, among other possible transformations,  spacetime isometries play a preferred role, owing to the physical importance of  linear and angular momenta; however, for  two opposite charges in the presence of a pure magnetic or pure electric field,   pseudo-momentum is conserved~\cite{drozelec} and a possible generalisation might be relevant.


   \end{document}